# Role-Based Access Controls

David F. Ferraiolo and D. Richard Kuhn
National Institute of Standards and Technology
Technology Administration
U.S. Department of Commerce
Gaithersburg, Md. 20899 USA

**ABSTRACT**

While Mandatory Access Controls (MAC) are appropriate for multilevel secure military applications, Discretionary Access Controls (DAC) are often perceived as meeting the security processing needs of industry and civilian government. This paper argues that reliance on DAC as the principal method of access control is unfounded and inappropriate for many commercial and civilian government organizations. The paper describes a type of non-discretionary access control - role-based access control (RBAC) - that is more central to the secure processing needs of non-military systems than DAC.

## 1      Introduction

The U.S. government has been involved in developing security technology for computer and communications systems for some time. Although advances have been great, it is generally perceived that the current state of security technology has, to some extent failed to address the needs of all. [1], [2] This is especially true of organizations outside the Department of Defense (DoD). [3]

The current set of security criteria, criteria interpretations, and guidelines has grown out of research and development efforts on the part of the DoD over a period of twenty plus years. Today the best known U.S. computer security standard is the Trusted Computer System Evaluation Criteria (TCSEC [4] ). It contains security features and assurances, exclusively derived, engineered and rationalized based on DoD security policy, created to meet one major security objective - preventing the unauthorized observation of classified information. The result is a collection of security products that do not fully address security issues as they pertain to unclassified sensitive processing environments. Although existing security mechanisms have been partially successful in promoting security solutions outside of the DoD [2] , in many instances these controls are less then perfect, and are used in lieu of a more appropriate set of controls.

The TCSEC specifies two types of access controls: Discretionary Access Controls (DAC) and Mandatory Access Controls (MAC). Since the TCSEC's appearance in December of 1983, DAC requirements have been perceived as being technically correct for commercial and civilian government security needs, as well as for single-level military systems. MAC is used for multi-level secure military systems, but its use in other applications is rare. The premise of this paper is that there exists a control, referred to as Role-Based Access Control (RBAC), that can be more appropriate and central to the secure processing needs within industry and civilian government than that of DAC, although the need for DAC will continue to exist.

## 2    Aspects of Security Policies

Recently, considerable attention has been paid to researching and addressing the security needs of commercial and civilian government organizations. It is apparent that significant and broad sweeping security requirements exist outside the Department of Defense. [2] , [5] , [6] Civilian government and corporations also rely heavily on information processing systems to meet their individual operational, financial, and information technology requirements. The integrity, availability, and confidentiality of key software systems, databases, and data networks are major concerns throughout all sectors. The corruption, unauthorized disclosure, or theft of corporate resources could disrupt an organization's operations and have immediate, serious financial, legal, human safety, personal privacy and public confidence impact.

Like DoD agencies, civilian government and commercial firms are very much concerned with protecting the confidentiality of information. This includes the protection of personnel data, marketing plans, product announcements, formulas, manufacturing and development techniques. But many of these organizations have even greater concern for integrity. [1]

Within industry and civilian government, integrity deals with broader issues of security than confidentiality. Integrity is particularly relevant to such applications as funds transfer, clinical medicine, environmental research, air traffic control, and avionics. The importance of integrity concerns in defense systems has also been studied in recent years. [7] , [8]

A wide gamut of security policies and needs exist within civilian government and private organizations. An organizational meaning of security cannot be presupposed. Each organization has unique security requirements, many of which are difficult to meet using traditional MAC and DAC controls.

As defined in the TCSEC and commonly implemented, DAC is an access control mechanism that permits system users to allow or disallow other users access to objects under their control:

A means of restricting access to objects based on the identity of subjects and/or groups to which they belong. The controls are discretionary in the sense that a subject with a certain access permission is capable of passing that permission (perhaps indirectly) on to any other subject (unless restrained by mandatory access control). [4]
DAC, as the name implies, permits the granting and revoking of access privileges to be left to the discretion of the individual users. A DAC mechanism allows users to grant or revoke access to any of the objects under their control without the intercession of a system administrator.

In many organizations, the end users do not ``own'' the information for which they are allowed access. For these organizations, the corporation or agency is the actual ``owner'' of system objects as well as the programs that process it. Control is often based on employee functions rather than data ownership.

Access control decisions are often determined by the roles individual users take on as part of an organization. This includes the specification of duties, responsibilities, and qualifications. For example, the roles an individual associated with a hospital can assume include doctor, nurse, clinician, and pharmacist. Roles in a bank include teller, loan officer, and accountant. Roles can also apply to military systems; for example, target analyst, situation analyst, and traffic analyst are common roles in tactical systems. A role based access control (RBAC) policy bases access control decisions on the functions a user is allowed to perform within an organization. The users cannot pass access permissions on to other users at their discretion. This is a fundamental difference between RBAC and DAC.

Security objectives often support a higher level organizational policy, such as maintaining and enforcing the ethics associated with a judge's chambers, or the laws and respect for privacy associated with the diagnosis of ailments, treatment of disease, and the administering of medicine with a hospital. To support such policies, a capability to centrally control and maintain access rights is required. The security administrator is responsible for enforcing policy and represents the organization.

The determination of membership and the allocation of transactions to a role is not so much in accordance with discretionary decisions on the part of a system administrator, but rather in compliance with organization-specific protection guidelines. These policies are derived from existing laws, ethics, regulations, or generally accepted practices. These policies are non-discretionary in the sense that they are unavoidably imposed on all users. For example, a doctor can be provided with the transaction to prescribe medicine, but does not possess the authority to pass that transaction on to a nurse.

RBAC is in fact a form of mandatory access control, but it is not based on multilevel security requirements. As defined in the TCSEC, MAC is

A means of restricting access to objects based on the sensitivity (as represented by a label) of the information contained in the objects and the formal authorization (i.e. clearance) of subjects to access information of such sensitivity. [4]
Role based access control, in many applications (e.g. [9] , [10] , [11] is concerned more with access to functions and information than strictly with access to information.

The act of granting membership and specifying transactions for a role is loosely analogous to the process of clearing users (granting membership) and the labeling (associate operational sensitivities) of objects within the DoD. The military policy is with respect to one type of capability: who can read what information. For these systems the unauthorized flow of information from a high level to a low level is the principal concern. As such, constraints on both reads and writes are in support of that rule. Within a role-based system, the principal concern is protecting the integrity of information: ``who can perform what acts on what information.''

A role can be thought of as a set of transactions that a user or set of users can perform within the context of an organization. Transactions are allocated to roles by a system administrator. Such transactions include the ability for a doctor to enter a diagnosis, prescribe medication, and add a entry to (not simply modify) a record of treatments performed on a patient. The role of a pharmacist includes the transactions to dispense but not prescribe prescription drugs. Membership in a role is also granted and revoked by a system administrator.

Roles are group oriented. For each role, a set of transactions allocated the role is maintained. A transaction can be thought of as a transformation procedure [1] (a program or portion of a program) plus a set of associated data items. In addition, each role has an associated set of individual members. As a result, RBACs provide a means of naming and describing many-to-many relationships between individuals and rights. Figure 1 depicts the relationships between individual users, roles/groups, transformation procedures, and system objects.

The term transaction is used in this paper as a convenience to refer to a binding of transformation procedure and data storage access. This is not unlike conventional usage of the term in commercial systems. For example, a savings deposit transaction is a procedure that updates a savings database and transaction file. A transaction may also be quite general, e.g. ``read savings file''. Note however, that ``read'' is not a transaction in the sense used here, because the read is not bound to a particular data item, as ``read savings file'' is.

The importance of control over transactions, as opposed to simple read and write access, can be seen by considering typical banking transactions. Tellers may execute a savings deposit transaction, requiring read and write access to specific fields within a savings file and a transaction log file. An accounting supervisor may be able to execute correction transactions, requiring exactly the same read and write access to the same files as the teller. The difference is the process executed and the values written to the transaction log file.

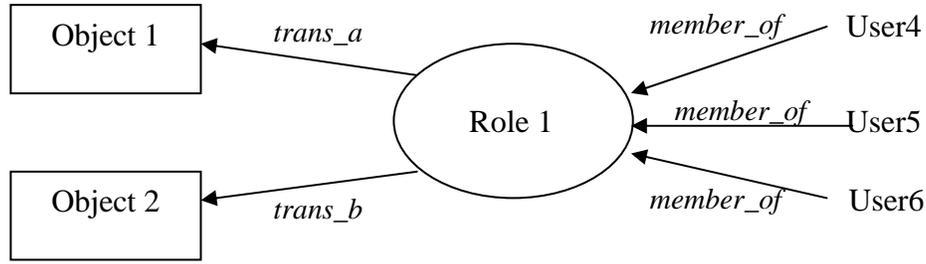

**Figure 1:** Role relationships

The applicability of RBAC to commercial systems is apparent from its widespread use. Baldwin [9] describes a database system using roles to control access. Nash and Poland [10] discuss the application of role based access control to cryptographic authentication devices commonly used in the banking industry. Working with industry groups, the National Institute of Standards and Technology has developed a proposed standard, ``Security Requirements for Cryptographic Modules,'' (Federal Information Processing Standard 140-1) [11] that will require support for access control and administration through roles. To date, these role based systems have been developed by a variety of organizations, with no commonly agreed upon definition or recognition in formal standards. *Role based access controls described in this paper address security primarily for application-level systems, as opposed to general purpose operating systems.*

## 3    Formal Description of RBAC

To clarify the notions presented in the previous section, we give a simple formal description, in terms of sets and relations, of role based access control. No particular implementation mechanism is implied.

For each subject, the active role is the one that the subject is currently using:

   $AR(s: subject) = \{the\ active\ role\ for\ subject\ s\}.$

Each subject may be authorized to perform one or more roles:

   $RA(s: subject) = \{authorized\ roles\ for\ subject\ s\}.$

Each role may be authorized to perform one or more transactions:

   $TA(r: role) = \{transactions\ authorized\ for\ role\ r\}.$

Subjects may execute transactions. The predicate *exec(s,t)* is true if subject *s* can execute transaction *t* at the current time, otherwise it is false:

   $exec(s: subject, t: tran) = true\ iff\ subject\ s\ can\ execute\ transaction\ t.$

Three basic rules are required:

1. Role assignment: A subject can execute a transaction only if the subject has selected or been assigned a role:

   $$\forall s : subject, t : tran, (exec(s,t) \Rightarrow AR(s) \neq \emptyset).$$

   The identification and authentication process (e.g. login) is not considered a transaction. All other user activities on the system are conducted through transactions. Thus all active users are required to have some active role.

2. Role authorization: A subject's active role must be authorized for the subject:

   $$\forall s : subject, (AR(s) \subseteq RA(s)).$$

   With (1) above, this rule ensures that users can take on only roles for which they are authorized.

3. Transaction authorization: A subject can execute a transaction only if the transaction is authorized for the subject's active role:

   $$\forall s : subject, t : tran, (exec(s,t) \Rightarrow t \in TA(AR(s))).$$

With (1) and (2), this rule ensures that users can execute only transactions for which they are authorized. Note that, because the conditional is ``only if'', this rule allows the possibility that additional restrictions may be placed on transaction execution. That is, the rule does not guarantee a transaction to be executable just because it is in *TA(AR(s))*, the set of transactions potentially executable by the subject's active role. For example, a trainee for a supervisory role may be assigned the role of ``Supervisor'', but have restrictions applied to his or her user role that limit accessible transactions to a subset of those normally allowed for the Supervisor role.

In the preceding discussion, a transaction has been defined as a transformation procedure, plus set of data items accessed by the transformation procedure. Access control in the rules above does not require any checks on the user's right to access a data object, or on the transformation procedure's right to access a data item, since the data accesses are built into the transaction. Security issues are addressed by binding operations and data into a transaction at design time, such as when privacy issues are addressed in an insurance query transaction.

It is also possible to redefine the meaning of ``transaction'' in the above rules to refer only to the transformation procedure, without including a binding to objects. This would require a fourth rule to enforce control over the modes in which users can access objects through transaction programs. For example, a fourth rule such as

4. $\forall s : subject, t : tran, o : object, (exec(s,t) \Rightarrow access(AR(s),t,o,x)).$

could be defined using a transaction (redefined to transformation procedure) to object access function *access(r, i, o, x)* which indicates if it is permissible for a subject in role *r* to access object *o* in mode *x* using transaction *t*, where *x* is taken from some set of modes such as read, write, append. Note that the Clark-Wilson access control triple could be implemented by letting the modes *x* be the access modes required by transaction *t*, and having a one-to-one relationship between subjects and roles. RBAC, as presented in this paper, thus includes Clark and Wilson access control as a special case.

Use of this fourth rule might be appropriate, for example, in a hospital setting. A doctor could be provided with read/write access to a prescription file, while the hospital pharmacist might have only read access. (Recall that use of the first three rules alone requires binding the transaction program *t* and data objects that *t* can access, and only controls access to the transactions.) This alternative approach using the fourth rule might be helpful in enforcing confidentiality requirements.

Another use of RBAC is to support integrity. Integrity has been defined in a variety of ways, but one aspect [8] of integrity is a requirement that data and processes be modified only in authorized ways by authorized users. This seems to be a reasonable security objective for many real systems, and RBAC should be applicable to such systems.

In general, the problem of determining whether data have been modified only in authorized ways can be as complex as the transaction that did the modification. For this reason, the practical approach is for transactions to be certified and trusted. If transactions must be trusted then access control can be incorporated directly into each transaction. Requiring the system to control access of transaction programs to objects through the access function used in rule (4) might then be a useful form of redundancy, but it could involve significant overhead for a limited benefit in enforcing integrity requirements. Therefore, inclusion of a transaction to object access control function in RBAC would be useful in some, but not all applications.

## 4      Centrally Administering Security Using RBAC

RBAC is flexible in that it can take on organizational characteristics in terms of policy and structure. One of RBAC's greatest virtues is the administrative capabilities it supports.

Once the transactions of a Role are established within a system, these transactions tend to remain relatively constant or change slowly over time. The administrative task consists of granting and revoking membership to the set of specified named roles within the system. When a new person enters the organization, the administrator simply grants membership to an existing role. When a person's function changes within the organization, the user membership to his existing roles can be easily deleted and new ones granted. Finally, when a person leaves the organization, all memberships to all Roles are deleted. For an organization that experiences a large turnover of personnel, a role-based security policy is the only logical choice.

In addition, roles can be composed of roles. For example, a Healer within a hospital can be composed of the roles Healer, Intern, and Doctor. Figure 2 depicts an example of such a relationship.

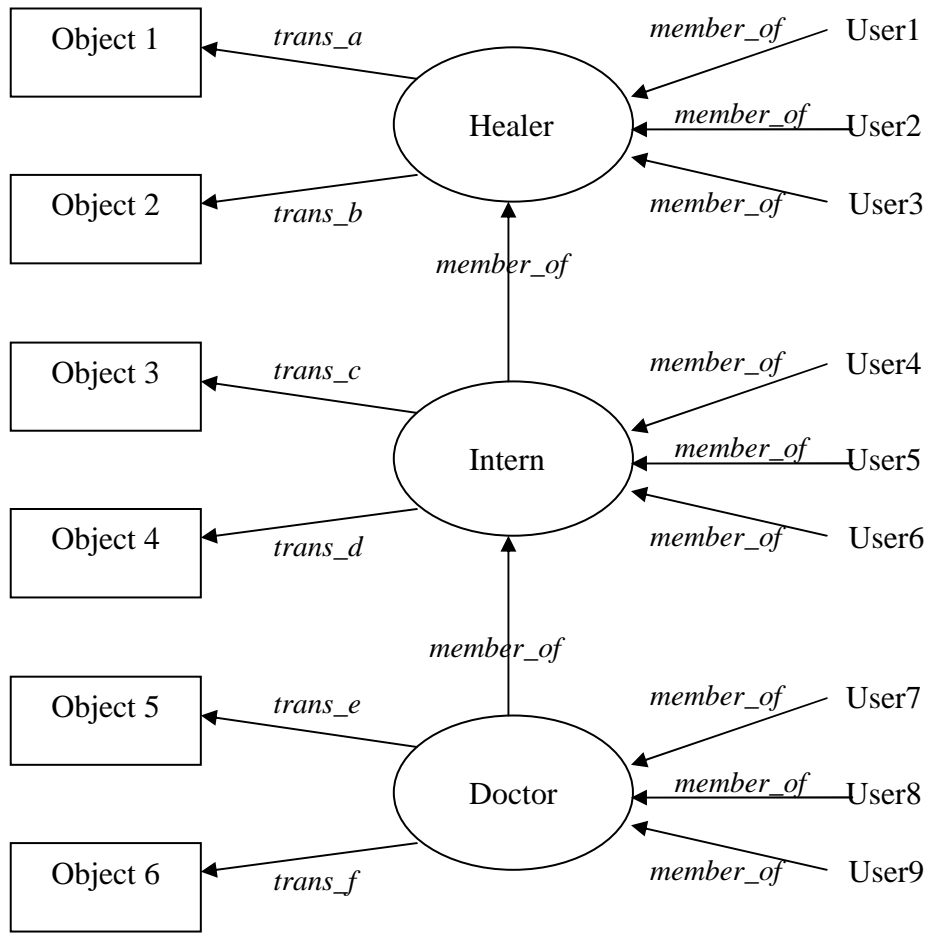

**Figure 2:** Multi-Role relationships

By granting membership to the Role Doctor, it implies access to all transactions defined by Intern and Healer, as well as those of a Doctor. On the other hand, by granting membership to the Intern role, this implies transactions of the Intern and Healer not the Doctor. However, by granting membership to the Healer role, this only allows access to those resources allowed under the role Healer.

# 5 Principle of Least Privilege

The principle of least privilege has been described as important for meeting integrity objectives. [8] The principle of least privilege requires that a user be given no more privilege than necessary to perform a job. Ensuring least privilege requires identifying what the user's job is, determining the minimum set of privileges required to perform that job, and restricting the user to a domain with those privileges and nothing more. By denying to subjects transactions that are not necessary for the performance of their duties, those denied privileges cannot be used to circumvent the organizational security policy. Although the concept of least privilege currently exists within the context of the TCSEC, requirements restrict those privileges of the system administrator. Through the use of RBAC, enforced minimum privileges for general system users can be easily achieved.

# 6 Separation of Duties

RBAC mechanisms can be used by a system administrator in enforcing a policy of separation of duties. Separation of duties is considered valuable in deterring fraud since fraud can occur if an opportunity exists for collaboration between various job related capabilities. Separation of duty requires that for particular sets of transactions, no single individual be allowed to execute all transactions within the set. The most commonly used examples are the separate transactions needed to initiate a payment and to authorize a payment. No single individual should be capable of executing both transactions. Separation of duty is an important consideration in real systems. [1] , [12] , [13] , [14] The sets in question will vary depending on the application. In real situations, only certain transactions need to be restricted under separation of duty requirements. For example, we would expect a transaction for ``authorize payment'' to be restricted, but a transaction ``submit suggestion to administrator'' would not be.

Separation of duty can be either static or dynamic. Compliance with static separation requirements can be determined simply by the assignment of individuals to roles and allocation of transactions to roles. The more difficult case is dynamic separation of duty where compliance with requirements can only be determined during system operation. The objective behind dynamic separation of duty is to allow more flexibility in operations. Consider the case of initiating and authorizing payments. A static policy could require that no individual who can serve as payment initiator could also serve as payment authorizer. This could be implemented by ensuring that no one who can perform the initiator role could also perform the authorizer role. Such a policy may be too rigid for commercial use, making the cost of security greater than the loss that might be expected without the security. More flexibility could be allowed by a dynamic policy that allows the same individual to take on both initiator and authorizer roles, with the exception that no one could authorize payments that he or she had initiated. The static policy could be implemented by checking only roles of users; for the dynamic case, the system must use both role and user ID in checking access to transactions.

Separation of duty is necessarily determined by conditions external to the computer system. The Clark-Wilson [1] scheme includes the requirement that the system maintain the separation of duty requirement expressed in the access control triples. Enforcement is on a per-user basis, using the user ID from the access control triple. As discussed above, user functions can be conveniently separated by role, since many users in an organization typically perform the same function and have the same access rights on TPs and data. Allocating access rights according to role is also helpful in defining separation of duty in a way that can be enforced by the system.

# 7    Summary and Conclusions

In many organizations in industry and civilian government, the end users do not ``own'' the information for which they are allowed access. For these organizations, the corporation or agency is the actual ``owner'' of system objects, and discretionary access control may not be appropriate. Role-Based Access Control (RBAC) is a nondiscretionary access control mechanism which allows and promotes the central administration of an organizational specific security policy.

Access control decisions are often based on the roles individual users take on as part of an organization. A role specifies a set of transactions that a user or set of users can perform within the context of an organization. RBAC provide a means of naming and describing relationships between individuals and rights, providing a method of meeting the secure processing needs of many commercial and civilian government organizations.

Various forms of role based access control have been described and some are used in commercial systems today, but there is no commonly accepted definition or formal standards encompassing RBAC. As such, evaluation and testing programs for these systems have not been established as they have for systems conforming to the Trusted Computer Security Evaluation Criteria. This paper proposed a definition of The requirements and access control rules for RBAC proposed in this paper could be used as the basis for a common definition of access controls based on user roles.